\documentstyle[12pt,psfig]{article}
\newcommand{\be}{\begin{equation}}
\newcommand{\ee}{\end{equation}}

\begin{document}
\hspace*{\fill} LMU--TPW--98/01, MPI--PhT--98/01\\[3ex]

\begin{center}
\Large\bf
Cosmological implications of a light dilaton\\[4ex]
\normalsize \rm
R.\ Dick$^{a\ast}$ and M.\ Gaul$^{a,b\mbox{\dag}}$
 \\[1ex] {\small\it
 ${}^a$Sektion Physik der Ludwig--Maximilians--Universit\"at\\
Theresienstr.\ 37, 80333 M\"unchen,
Germany\\[1ex] 
 ${}^b$Max--Planck--Institut f\"ur Physik \\
 F\"ohringer Ring 6, 80805 M\"unchen, Germany}
\end{center}

\vspace{5ex}
\noindent
{\bf Abstract}: Supersymmetric Peccei--Quinn symmetry and string theory predict
a complex scalar field comprising a dilaton and an axion. These
fields are massless at high energies, but
it is known since long that the axion is stabilized in an instanton dominated vacuum.
Instantons and axions together also provide a mechanism to stabilize a dilaton, 
thus accounting for a dilaton as a possible cold dark matter
component accompanying the axion. We briefly review the 
prospects of this scenario and point out further implications.

\vspace{\fill}
\noindent
------------------------------------------------------------\\
{\footnotesize Based on a talk at COSMO--97, First International Workshop
on Particle Physics and the Early Universe, Ambleside (England) 15--19 September
1997.\\
 ${}^\ast$Rainer.Dick@physik.uni--muenchen.de, ${}^{\mbox{\dag}}$mred@mppmu.mpg.de.}

\newpage
Cosmology and physics of the early universe play an 
increasing role in physics, nurtured by seminal 
experimental and theoretical interactions
with particle physics and astrophysics.
A particularly exciting field of interaction between particle physics,
astrophysics and cosmology concerns the nature of the dark matter in the universe:
While there is consensus that an appreciable fraction has to
consist of non-baryonic components, it is unclear yet whether we are talking
about LSPs or axions eventually coming with a dilaton or something else,
and what the contribution from massive neutrinos is. It is also unclear,
whether cold dark matter comes as WIMPs or through coherent oscillations
of very light (pseudo-)scalars. A light axion/dilaton pair motivated from
supersymmetric Peccei--Quinn symmetry or string theory is an interesting
candidate for the latter scenario.

The axion \cite{PQWW,inv} is widely recognized as
a leading competitor for the role of cold dark matter in the universe \cite{PWW,mst},
and intense searches are under way \cite{bibber}. However, in models which
originate from a supersymmetric theory or realize a duality symmetry
on an abelian subgroup, the axion $a$ is accompanied by a dilaton $\phi$,
which would also contribute an appreciable amount to the energy density 
of the universe. 
The relevant couplings before taking into account non-perturbative 
and curvature effects
are contained in a Lagrangian
\[
{\cal L}=-\frac{1}{2}
\partial_\mu\phi\cdot\partial^\mu\phi
-\frac{1}{2}\exp\!\Big(-2\frac{\phi}{f_\phi}\Big)
\partial_\mu a\cdot\partial^\mu a
-\frac{1}{4}\exp\!\Big(\frac{\phi}{f_\phi}\Big) F_{\mu\nu}{}^j F^{\mu\nu}{}_j +
\frac{a}{4f_\phi} {\tilde F}_{\mu\nu}{}^j F^{\mu\nu}{}_j,
\]
where we specialized to the supersymmetric/S--dual case \cite{stw}
in fixing the axion scale $f_a=(\alpha_s/2\pi)f_\phi$.

The observation that a CDM axion should come with a dilaton 
is already implicit in \cite{ENQ}, since it was pointed out
that a dilaton with a large coupling scale $f_\phi$ should also meet
cosmological constraints from $\Omega\leq 1$. 
It was shown in \cite{CCQR} that dilatons with $f_\phi$
of the order of the Planck mass should acquire a mass of the order
of the gravitino mass if supersymmetry is softly broken at low energies.
Another mass generating mechanism was proposed
in \cite{rddil}, where 
it was pointed out that instantons and axions together may generate
a dilaton mass due to the specific couplings of the dilaton to the
kinetic terms of the axion and the gluons. This implies a scenario where
a dilaton is stabilized at a very late stage together with an axion, whence
both particles would arise through coherent oscillations at the chiral
phase transition. On the other hand, dilatons which are stabilized 
at a high temperature generically acquire higher masses
and may appear as WIMPs rather than through coherent oscillations.
In this sense dilatons were mentioned as dark matter already in \cite{GV,DV},
and mechanisms for dilaton stabilization at or above the
SUSY scale are
discussed in \cite{DIN,CCQR,DP,BD}.

\begin{figure}[ht]
\centerline{\psfig{figure=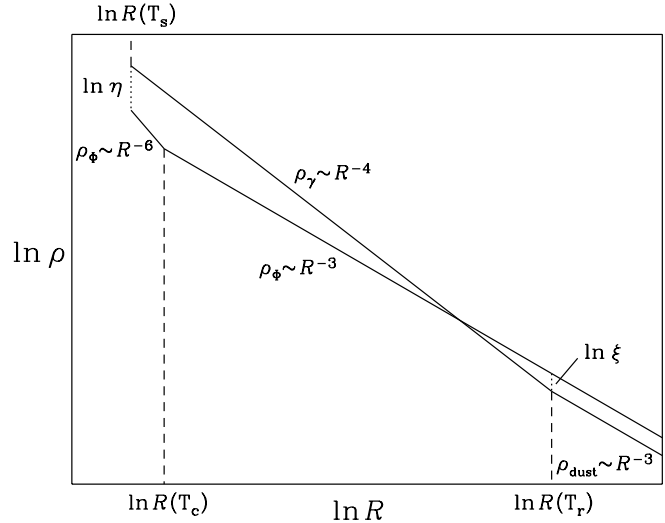,width=9cm}}
\vspace{-0.5cm}\caption{Transition from a stiff fluid to CDM. \label{fig1}}
\end{figure}

In the present contribution we are mainly concerned with the prospects
of a dilaton as a CDM component, neglecting interesting implications
of possible dilaton--curvature couplings, for which 
we refer to \cite{GV,DP,bradi} and references there.
However, before focussing
on axions and dilatons let us make a general remark on cold dark matter
from particle physics beyond the standard model:

The characteristic features of cold dark matter 
are very weak coupling and negligible pressure,
whence the energy $\rho_\phi$
in the CDM degrees of freedom decays with
the third power of the scale factor {\sl R}${}^{-3}$ and catches up with
the radiation dominated heat reservoir according 
to $\rho_\phi/\rho_\gamma\sim \mbox{\sl R}\sim\sqrt{t}$. Due to its weak
coupling CDM is also decoupled from the heat bath,
and this has simple but interesting implications: Both WIMPs and
light axions and dilatons begin to behave like dust when mass terms are created
during a phase transition or become comparable to the temperature.
However, before the onset of mass terms or for temperatures well above the
mass scale of a WIMP, CDM components can be considered as factually or effectively
massless, and the question arises about evolution of their energy density
during this early phase: If the CDM degrees of freedom, in spite of their
weak coupling, have a sufficiently strong internal coupling for internal
thermalization, then their energy decays in parallel to the 
heat bath $\rho_\phi\sim\mbox{\sl R}^{-4}$ and $\rho_\phi/\rho_\gamma$
remains constant during the pre-CDM phase.
However, if all the couplings of the CDM components are very weak,
then internal thermalization cannot be assumed, and before onset of mass terms
CDM degrees of freedom behave like a stiff fluid $p_\phi=\rho_\phi$,
resulting in a sharp decline of their energy 
density $\rho_\phi/\rho_\gamma\sim \mbox{\sl R}^{-2}$
above the temperature $T_c$ where mass terms become relevant
(see Fig.\ \ref{fig1}).

Superficially, we refer to the degrees of freedom in the stiff fluid as moduli.
In Fig.\ \ref{fig1} $T_s$ labels a scale where both the moduli and the
radiation originate (this might be a string scale, but this we do not know). 
The scale $T_c$ indicates, that  a transition
from the stiff fluid behaviour to CDM--like behaviour emerges upon
creation or increasing importance of mass terms, and $T_r$ is
the scale of photon recombination, when most energy in the heat reservoir
is converted into pressureless matter. 
Here, all temperatures refer to the radiation dominated heat bath,
whence $T\sim\mbox{\sl R}^{-1}$.

The scale $T_c$ is readily calculated in terms
of $T_s$, $T_r$ and the 
ratios $\eta=\rho_{\gamma}(T_s)/\rho_{\phi}(T_s)$
and $\xi=\rho_{\phi}(T_r)/\rho_{\gamma}(T_r)$:
\be\label{tc}
T_c=(\eta\xi T_r T_s^2)^{1/3}.
\ee
The corresponding time scales can be inferred from the Friedmann
equation
under the proviso of negligible coupling between the radiation and the
moduli: For $T>T_c$ the scale factor evolves
according to
\[
\frac{2}{\sqrt{3}m_{Pl}}\rho_{0\gamma}(t-t_0)=
x\sqrt{x^2\rho_{0\gamma}+\rho_{0\phi}}
-\sqrt{\rho_{0\gamma}+\rho_{0\phi}}
\]
\[
-
\frac{\rho_{0\phi}}{\sqrt{\rho_{0\gamma}}}
\ln\!\Bigg(
\frac{x\sqrt{\rho_{0\gamma}}+\sqrt{x^2\rho_{0\gamma}+\rho_{0\phi}}}
{\sqrt{\rho_{0\gamma}}+\sqrt{\rho_{0\gamma}+\rho_{0\phi}}}
\Bigg), 
\]
while for lower temperatures the evolution follows
\[
\frac{\sqrt{3}}{2m_{Pl}}\rho_{0\phi}^2(t-t_0)=(x\rho_{0\phi}-2\rho_{0\gamma})
\sqrt{x\rho_{0\phi}+\rho_{0\gamma}}
-(\rho_{0\phi}-2\rho_{0\gamma})
\sqrt{\rho_{0\phi}+\rho_{0\gamma}}.
\]

Here $x={\mbox{\sl R}}/{\mbox{\sl R}}_0$, $m_{Pl}=2.4\times 10^{18}$ GeV,
and the subscript 0 in each equation
indicates an arbitrary
fiducial time during the respective epoch.

Consider the heterotic string \cite{GHMR}, e.g.: This predicts an effective
number of massless states $g_{\ast}=7560$ below the string scale, among which 120
belong to the very weakly coupled
gravitational sector. Assuming thermalization of the other degrees of freedom 
and equipartition at the string scale
gives an estimate $\eta=62$. Furthermore, it was pointed out in \cite{rdscale} that
the heterotic string scale should be as low as $T_s\simeq 4\times 10^{16}$ GeV, 
because for 
an earlier transition
the product of comoving time and string scale would spoil
the uncertainty relation. Inserting $T_r\simeq 0.3$ eV and $\xi\simeq 10$ in (\ref{tc}) then gives
 $T_c\sim 10^9$ GeV, while the same calculation with $T_s=m_{Pl}$ 
yields $T_c\sim 10^{10}$ GeV.
 These values comply nicely with
bounds on an axion scale
(see e.g.\ \cite{georg}), and with Pati's proposal of a grand fiesta of new physics 
around $10^{11}$ GeV \cite{jogesh}.

On the other hand, if the moduli would be thermalized already for temperatures
above $T_c$, eq.\ (\ref{tc}) would be replaced by $T_c=\eta\xi T_r$, and large values of
 $T_c$ would require a huge ratio $\eta$.

Eq.\ (\ref{tc}) generically indicates a high temperature phase transition if we insist on
continuous evolution of the energy density in the moduli. Conversely, it means
that the energy in an axion and a dilaton
stabilized through instanton effects at a scale $T\sim 1$ GeV arises from conversion
of a tiny fraction $\sim 10^{-8}\rho_\gamma$ of radiation 
during the chiral phase transition.

The instanton--induced potential \cite{rddil}
\be\label{axipot}
V(a,\phi)=\frac{1}{6}m_\phi^2 f_\phi^2\Big(2\exp\!\Big(\frac{\phi}{f_\phi}\Big)
+\exp\!\Big(-2\frac{\phi}{f_\phi}\Big)\Big)+m_a^2 f_a^2\Big(1-\cos\!\Big(\frac{a}{f_a}\Big)\Big)
\ee
yields parameters $m_\phi f_\phi\simeq m_a f_a\simeq m_\pi f_\pi$.
However, a dilaton mass raises the same issues as an axion mass:
A large--$f_\phi$ dilaton in an open or flat universe would prevent the chiral
phase transition in the same way
as an axion \cite{PWW}, 
and the well-known upper bound on decay constants
arising from chiral symmetry breaking
and $\Omega\leq 1$ is $f\leq 10^{12}$ GeV.
Possibilities to relax this bound include
large entropy production \cite{LSSS} or a 
subsequent phase of 
accelerated expansion \cite{linde}. Recently, Banks and Dine
also proposed two very different scenarios, supposing
either moduli domination
until a reheating temperature between the QCD scale and the scale
of nucleosynthesis is reached,
or an axion driven to $\sqrt{\langle a^2\rangle}\leq 10^{-4}f_a$
by an effective potential during inflation \cite{BD}.
However, overall the bound $f\leq 10^{12}$ GeV
is very robust, and even
the assumption of strong coupling
for temperatures
well above the 
phase transition turned 
out not to ameliorate the problem \cite{CKK}.

Here we would like to briefly comment on another proposal:
Above the phase transition both the axion and the dilaton are massless
and one should expect strong fluctuations in both fields.
Therefore, one might expect chiral symmetry breaking also
to occur in presence
of a large--$f$ axion, but not everywhere at the same time, i.e.\
the transition hypersurfaces would not be flat (for $k=0$)
or hyperbolic (for $k=-1$). 
As a consequence, a large--$f$ axidilaton would induce an extra time
scale and generate extra perturbations
during the transition. Of course, there still would exist an upper bound, 
since very large $f$ would imply too long a time
scale for the transition and generate too large perturbations at the QCD scale.
Stated differently, 
the average displacement of the axidilaton from the vacua
of the potential (\ref{axipot}) prevents emergence of the chiral symmetry
breaking low temperature phase, since it costs too much energy
to create a mass term. But the scenario just mentioned presumes, that
this would not apply to those regions 
where the axidilaton field
is close to the minimum of (\ref{axipot}).
In those regions the instanton gas
could dominate the partition function 
and induce both a quark condensate and axidilaton mass terms,
thus confining the axion and the dilaton in that region to absolute values $\ll f$.
Outside of those
instanton dominated bubbles the axidilaton 
continues to fluctuate (almost) freely, until it comes close enough to zero,
whence it also gets trapped through the onset of
the phase transition.

The caveat with this proposal concerns the emergence 
of fluctuations required to relax the bound on $f$: 
The quantum fluctuations of the axidilaton
in the radiation dominated background are only vacuum fluctuations,
which are neglected e.g.\ in 
calculating the seeds for structure
formation in de Sitter space \cite{inf}, and would also be subtracted
if one calculates the two-point function according to Bunch and Davies or
Vilenkin and Ford \cite{qf1}. Therefore, any appreciable effect should result
from thermal fluctuations, and this touches upon the problem
of decoherence in subluminally expanding spacetimes.
We have nothing to say about the latter problem at this stage, and therefore we
conclude with the assertion that one can think of plausible scenarios
for accomodating chiral symmetry breaking in $\Omega\leq 1$ universes
with large--$f$ axions and dilatons, but overall the bound $f\leq 10^{12}$ GeV
seems very robust.\\[2ex]
{\bf Acknowledgement:} R.D.\ would like to thank the organizers of COSMO--97
for hospitality during a very interesting and stimulating meeting. 
Support by the DFG
through SFB 375--95 is gratefully acknowledged.


\end{document}